# A liquid nitrogen cooled superconducting transition edge sensor with ultra-high responsivity and GHz operation speeds

Paul Seifert[1†], José Ramón Durán Retamal[1†], Rafael Luque Merino[1], Hanan Herzig Sheinfux[1], John N. Moore[1], Mohammed Ali Aamir[1], Takashi Taniguchi[2], Kenji Wantanabe[3], Kazuo Kadowaki[4], Massimo Artiglia[5], Marco Romagnoli[5] and Dmitri K. Efetov[1]*

1. ICFO - Institut de Ciencies Fotoniques, The Barcelona Institute of Science and Technology, Castelldefels, Barcelona, 08860, Spain
2. International Center for Materials Nanoarchitectonics, National Institute for Materials Science, 1-1 Namiki, Tsukuba 305-0044, Japan
3. Research Center for Functional Materials, National Institute for Materials Science, 1-1 Namiki, Tsukuba 305-0044, Japan
4. Algae Biomass and Energy System (ABES) Research & Development Center & Institute of Materials Science University of Tsukuba 1-1-1, Tennodai, Tsukuba-shi, 305-8572, Japan
5. CNIT - National Interuniversity Consortium for Telecommunications, TeCIP -Institute of Communication, Information and Perception Technologies - Scuola Superiore Sant'Anna via Moruzzi, 1, 56124 Pisa, Italy

* E-mail: dmitri.efetov@icfo.eu, † equal contribution

**Photodetectors based on nano-structured superconducting thin films are currently some of the most sensitive quantum sensors and are key enabling technologies in such broad areas as quantum information, quantum computation and radio-astronomy. However, their broader use is held back by the low operation temperatures which require expensive cryostats. Here, we demonstrate a nitrogen cooled superconducting transition edge sensor, which shows orders of magnitude improved performance characteristics of any superconducting detector operated above 77K, with a responsivity of $9.61 \cdot 10^4$ V/W, noise equivalent power of 15.9 fW/Hz$^{-1/2}$ and operation speeds up to GHz frequencies. It is based on van der Waals heterostructures of the high temperature superconductor $Bi_2Sr_2CaCu_2O_{8+\delta}$, which are shaped into nano-wires with ultra-small form factor. To highlight the versatility of the detector we demonstrate its fabrication and operation on a telecom grade SiN waveguide chip. Our detector significantly relaxes the demands of practical applications of superconducting detectors and displays its huge potential for photonics based quantum applications.**

Because of their high detection efficiencies, low dark count rates and small noise equivalent powers (NEP), photo-detectors based on thin film superconductors (SCs) are crucial components in the development of revolutionary quantum communication protocols, such as quantum key distribution and the Bell inequality test[1–5]. Such outstanding performance is made possible by engineering superconducting nanostructures with ultra-low heat capacity, which use the steep SC transition edge to induce large resistance jumps by the heat generated from absorbed photons[1,2,6]. However, a broader use of SC detectors is held back by the low operation temperatures, which require expensive cryostats, and the highly disordered growth methods, which set an upper limit to the SC film thickness and impose strict restrictions to the employed SC materials and substrates. A breakthrough in this field would lie in the creation of ever thinner materials with ever higher superconducting transition temperatures ($T_c$), which would ultimately reach values beyond the boiling point of liquid nitrogen $T > 77K$.

On this path, cuprate high temperature superconductors (high-$T_c$), with their record-high $T_c > 100$K[7–9], are extremely promising. While their chemical sensitivity[10,11] and the difficulty in thin film preparation strongly limited the performance of previous cuprate based photodetectors[9,12–16], recent breakthroughs in fabrication protocols allowed the preparation of single crystals of the optimally doped cuprate compound $Bi_2Sr_2CaCu_2O_{8+\delta}$ (BSCCO), down to half a unit cell thick layers with thickness of $d \sim 1.5$nm and a $T_c \sim 91$K[17,18]. Belonging to the class of layered two-dimensional (2D) van der Waals (vdW) materials in which SC is confined to 2D planes[18], BSCCO can be mechanically exfoliated and non-invasively transferred onto any substrate. These entirely new capabilities also open previously unimaginable prospects for SC detectors, as BSCCO based detectors can be deterministically integrated into silicon photonics platforms.

Here we demonstrate a telecom wavelength high-$T_c$ superconducting transition edge sensor (TES) that can be operated above liquid nitrogen temperatures, with record responsivity and operation speeds, which are comparable to state-of-the-art commercial detectors, such as InGaAs photodiodes[9,12–16,19]. The detector is made from superconducting BSCCO and is enabled by a novel inert nanofabrication, which allows unprecedentedly small detector dimensions as compared to most previous high-$T_c$ detectors, with thickness $d \sim 15$nm and width $w \sim 100$nm (Figs. 1a-c). To highlight the material's versatility and integrability, we demonstrate operation of the BSCCO TES on a telecom grade SiN waveguide chip, which is laser illuminated through a grating coupler (Fig.1e and f). The photodetection principle of the BSCCO nano-wire is based on the sharp temperature dependent resistance at the superconducting transition edge, and temperature increase upon absorption of radiation, as is illustrated in Fig. 1d. The temperature increase upon absorption of radiation, locally drives the BSCCO into a resistive state which is stabilized via Joule self-heating and leads to a voltage spike across the device. At the same time, the superconducting surrounding of the nano-wire suppresses an electronic out-diffusion of heat, which is trapped at the thinnest part of the nano-structure[5].

The devices are prepared in inert argon atmosphere to avoid instantaneous oxidation of the BSCCO surface in air (SI Fig. S1). Few layer BSCCO sheets with thicknesses between 5 and 9 unit cells are mechanically exfoliated from bulk single crystals. These are then transferred onto the target substrate and encapsulated with thin sheets of insulating hexagonal boron-nitride layers (hBN), which prohibits further environmental degradation of the BSCCO. Electrical contacts to the BSCCO layer with contact resistances in the order of $R_c \sim 0.1$-$1$k$\Omega$ is established by overlapping BSCCO with pre-patterned gold electrodes on the target substrate (SI). Without destroying the hermetic seal of the hBN layer, the BSCCO crystals are then locally subjected to focused helium ion beam irradiation (He-FIB) (SI), which modifies metallic BSCCO regions into insulating ones, by means of defect formation and amorphization[20]. The scan ability of the He-FIB and the ultra-high definition of the so-created insulating BSCCO regions down to 10nm allow to create homogeneous superconducting BSCCO nano-wires and circuits with conducting channel widths of only $w \sim 100$nm (SI Fig. S3-S5).

Fig. 2a shows the temperature dependent resistance measurements $R$ of a 7 unit cell thin BSCCO crystal before (gray) and after (blue) nano-wire fabrication with $w \sim 100$nm. Strikingly, after FIB shaping the device shows a robust SC state, with an onset of superconductivity at $T \sim 90$K. However, the SC transition becomes broadened by inhomogeneities and doping induced by the FIB treatment, with the zero-resistance state setting in only at around $T \sim 40$K. Fig. 2b shows the two-terminal differential resistance $dV/dI$ measurements as a function of bias current $I_{bias}$ and temperature $T$. We find a sharp superconducting transition at the critical current

$I_c \sim 435\mu A$ at $T = 15K$, which is monotonically decreasing when $T$ is increased. In the superconducting state, contact resistance at the gold/BSCCO interface gives rise to a reproducible residual resistance with a peak around zero bias. Depending on the sweeping direction, $I_c$ develops a strongly pronounced hysteresis below $T \sim 30K$ which can be explained by Joule self-heating in the resistive state when the current is swept down (Fig. 2c and SI Fig. S9). Connecting the nano-wire in parallel to a 1kΩ resistor during photo-detection measurements (Fig. 1b), prohibits its latching into a self-heated resistive state after absorption of radiation.

We probe the photo response of the $I_{bias} \sim 0.1mA$ biased device with a telecom laser with wavelength $\lambda = 1550nm$. Fig. 3a displays photo response maps, obtained by scanning the normally incident laser spot across the device, as well as across the grating coupler of the SiN waveguide. The responsivity of the detector is defined as the measured photo-voltage divided by the incident laser power $V_{photo}/P_{laser}$. It exhibits a pronounced maximum when the laser is focused directly onto the nano-wire itself. For $V_{photo}/P_{laser}$ under free space coupling we assume full absorption of the laser power $P_{laser}$, which is incident on the metallic BSCCO nano-wire. Alternatively, the laser light can be injected into the waveguide, which then couples to the device through its evanescent mode and gives a sizeable photo-voltage response. Non-optimized grating coupling and propagation losses, result in a reduced overall efficiency (SI Fig. S11). However, recent studies of 2D materials on optimized photonic structures like waveguides[21,22] photonic crystals[23–25], ring resonators[26] and Fabry-Pérot microcavities[27], have demonstrated that light absorption can be enhanced to almost 100%.

We highlight the strong dependence of $V_{photo}$ on the superconducting transition edge, which shows a pronounced maximum at $I_c$. Fig. 3b compares the temperature dependent I/V characteristics (bottom) with the normalized photo-response $V_{photo}/V_{max}$ (top) under laser illumination with $P_{laser} = 100pW$ (top), where $V_{max}$ is the maximal response at $I_c$. The SC transition edge is sharply defined at low temperatures and gives rise to a large and sharp $V_{photo}$. As temperature is increased, both gradually broaden, but strikingly even above $T = 77K$, superconductivity is well formed and $V_{photo}$ shows a pronounced maximum and remains high. Fig. 3c shows the extracted maximum responsivity $V_{photo}/P_{laser}$ (blue) and the corresponding NEP (pink) as a function of device temperature $T$, where the NEP is obtained from the ratio of the theoretical thermal noise dominated voltage noise spectral density and $V_{photo}$[28]. At $T=15K$, the detector performance reaches an ultra-high responsivity of $V_{photo}/P_{laser} = 2.33 \cdot 10^7$ V/W resulting in an ultra-low $NEP = 55.2$ aW/Hz$^{-1/2}$. Most strikingly however, at $T = 77K$, the detector shows record performance for any SC detector, with a responsivity of $V_{photo}/P_{laser} = 9.61 \cdot 10^4$ V/W and a $NEP = 15.9$ fW/Hz$^{-1/2}$.

Finally, we investigate the intrinsic speed limits of the detector, by measuring its AC relaxation time $\tau$. To this end we employ a continuous wave photo-mixing technique, which was previously used to investigate the thermal relaxation in graphene (compare supporting information S12)[29]. The BSCCO nano-wire is irradiated with two interfering, frequency-detuned lasers, which form an optical beating pattern with oscillation frequency $\Delta f$ - the frequency difference of the two lasers[29]. Measurements of the AC photo-voltage $V_{mix}$ as a function of $\Delta f$ reveal a Lorentzian peak, which corresponds to an exponential decay in the time domain, and hence determines the detectors relaxation time $\tau$ (Fig. 4a). We extract two time scales, one ultra-fast component $\tau_1 \sim 220ps$ and a slower component $\tau_2 \sim 2ns$, which show only slight dependence on the sample temperature (Fig. 4b). These are longer than the thermal relaxation times found in bulk BSCCO crystals[30], and may correspond to suppressed heat diffusion mechanisms in these thin and nano-patterned devices. We show that these time-scales set the speed limits of the detector. At $T = 77K$ it can operate at frequencies up to $f \sim 100MHz$ without any sacrifice in responsivity, and can be even read out at GHz frequencies with a slightly reduced responsivity (Fig. 4c). Such ultra-fast operation speeds are absolutely unprecedented for a superconducting transition edge sensor, being many orders of magnitude faster than any previously reported TES[31,32].

Our study presents a breakthrough in the development of high-performance superconducting detectors and paves the way towards cheap, liquid nitrogen cooled quantum sensors. The presented unoptimized detector concept already shows record high performance characteristics, but it is feasible to push it to the ultimate limit of half a unit cell thickness. Alongside with the creation of more homogeneous nano-wire edges, this advancement will allow ever sharper SC transitions, and lead to the creation of liquid nitrogen cooled superconducting nano-wire single photon detectors (SNSPD) - a long standing goal in the field. The ease of integration into photonic circuits, will make such detectors a promising building block for the development of photonic quantum circuits.

Acknowledgements:
We are grateful for fruitful discussions with A. Montanaro, A. Santamato, M.A. Giambra and V. Sorianello and F .H. L. Koppens, and for technical support by J. Osmond. D.K.E. acknowledges support from the Ministry of Economy and Competitiveness of Spain through the "Severo Ochoa" program for Centres of Excellence in R&D (SE5-0522), Fundació Privada Cellex, Fundació Privada Mir-Puig, the Generalitat de Catalunya through the CERCA program, the H2020 Programme under grant agreement n° 820378, Project: 2D·SIPC and the La Caixa Foundation. K.W. and T.T. acknowledge support from the Elemental Strategy Initiative conducted by the MEXT, Japan, Grant Number JPMXP0112101001, JSPS KAKENHI Grant Number JP20H00354 and the CREST(JPMJCR15F3), JST. P. S acknowledges support from the Alexander-von-Humboldt Foundation and the German Federal Ministry for Education and Research through the Feodor-Lynen program.

Author contributions:
D.K.E. and P.S. conceived and designed the experiments; P.S., J.R.D. and R.L. performed the experiments; J.R.D., R.L.M. and P.S. fabricated the devices; P.S. and D.K.E. analyzed the data; P.S. and J.N.M. performed the theoretical modeling; H.H.S and M.A.A. provided technical support; K.K., T.T., K.W. contributed materials and M.A., M.R. contributed wave-guides; D.K.E. supported the experiments: D.K.E and P.S. wrote the paper.

**Supplementary Information** is available for this paper.

**Correspondence and requests for materials** should be addressed to D.K.E.

**Competing financial and non-Financial interests**:
The authors declare no competing financial and non-financial interests.

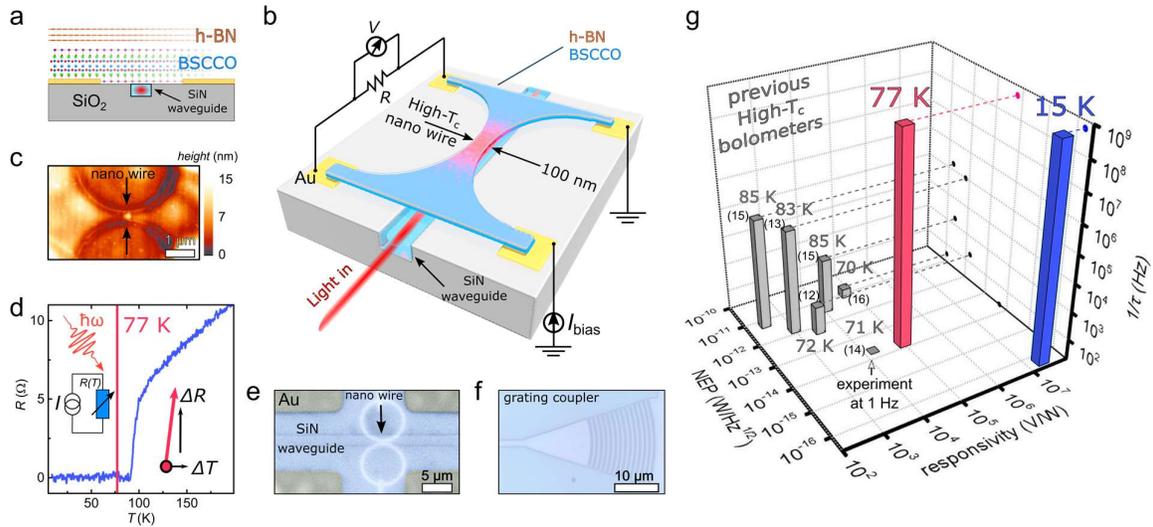

**FIG. 1. Detector concept. a** and **b**, Cross-section and schematic of the SiN waveguide coupled BSCCO superconducting transition edge sensor. An exfoliated 5-9 unit cell thin BSCCO crystal is transferred onto a SiN waveguide and pre-patterned gold electrodes, and covered with hexagonal boron nitride to prevent degradation. A nano-wire is patterned into the BSCCO at the position of the waveguide using focused helium ion beam lithography. **c**, AFM topography of the hBN/BSCCO nanowire. **d**, Sheet resistance of a pristine BSCCO device as a function of temperature with a $T_c = 91$K, above the boiling point of liquid nitrogen $T = 77$K. **e** and **f**, Optical image of the SiN waveguide integrated device and the grating coupler. **g,** Summary of responsivity, noise equivalent power (NEP) and operation frequency $1/\tau$ of the superconducting BSCCO detector at $T = 77$ K (pink bar) and $T = 15$ K (blue bar). For comparison the gray columns depict the performance of selected previous thermal detectors made from high-$T_c$ materials, including micro-bridges[12-14] and meander shaped devices[15,16]. For an extensive review on previous work we refer to[9].

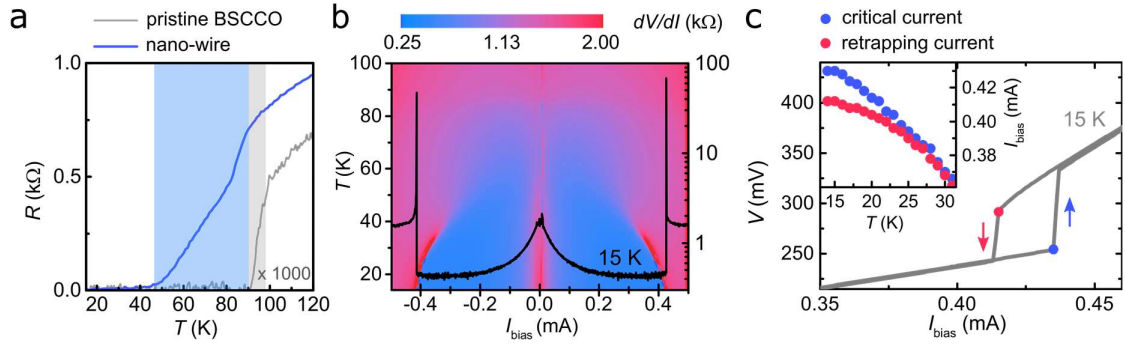

**FIG. 2. Electronic device properties. a**, Device resistance of the BSCCO before (gray) and after He-FIB nano-fabrication into wires of width $w \sim 100$nm (blue). The nano-wire shows a robust SC state with a broadened Sc transition. **b,** Differential resistance $dV/dI$ of the nano-wire device as a function of bias current $I_{bias}$ and temperature $T$, shows a sharp SC transition with Ic $\sim 435\mu A$ (15K) and $T_c \sim 91$K. The black line depicts a line cut at $T = 15$K. **c**, The current voltage characteristic $I/V$ of the nano-wire exhibit an open, temperature dependent (inset) hysteresis loop between up ($I_c$) and down (re-trapping current) swept $I_{bias}$.

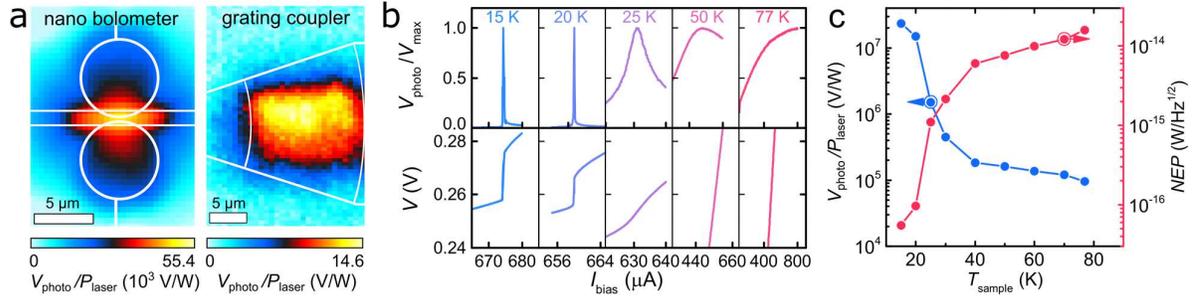

**FIG. 3. Optical response characteristics a**, Spatial map of the nanowire's optical responsivity under focused laser scanning at the position of the nanowire and at the position of the grating coupler at $T = 23$K (compare positions in Fig. 1e and f). **b,** Normalized photo-response $V_{photo}/V_{max}$ (top) and current-voltage characteristic $I/V$ (bottom) a function of bias current $I_{bias}$, measured at different temperatures $T$. **c,** Photo-voltage $V_{photo}/P_{laser}$ (blue) and calculated noise equivalent power (NEP, pink) as a function of device temperature $T$.

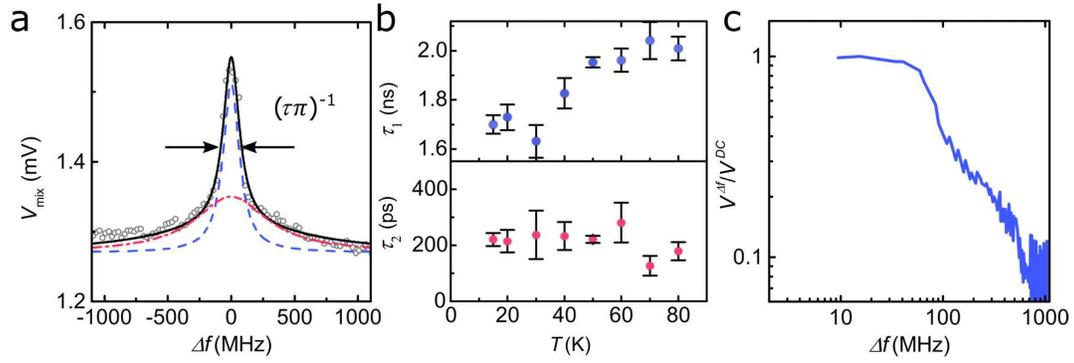

**FIG. 4. Frequency domain response from optical mixing. a**, AC photo-mixing voltage response $V_{mix}$ of the BSCCO nano-wire as a function of the difference frequency $\Delta f$ of the two interfering lasers at $T = 77$K. The fit (solid black line) is a double Lorentzian peak comprising a slow (dashed blue line) and a fast contribution (dash-dotted pink line) of width $1/\pi\tau$, corresponding to an exponential decay with relaxation time $\tau$. **b**, Extracted relaxation times $\tau_1$ and $\tau_2$ of the detector response as a function of device temperature. **c**, Normalized AC voltage response of the BSCCO detector at $T = 77$K at the difference frequency $\Delta f$.

**Methods**:

Materials and fabrication.
The hBN/BSCCO devices are assembled from micro-mechanically exfoliated crystals using a van der Waals assembly technique. BSCCO and hBN flakes are exfoliated onto a PDMS film on a glass substrate. We transfer the BSCCO onto pre-patterned Ti/Au electrodes on the SiN/SiO2 chip to make electrical contact. We subsequently cover the BSCCO flake with a suitable hBN crystal in order to avoid device degradation. The whole device fabrication is performed inside a glovebox under inert argon atmosphere.

For the fabrication of nano-wires, we locally irradiate the device with a 30keV helium-ion beam in a Zeiss Orion NanoFab. For patterning the structures we irradiate the BSCCO/hBN structure with ion doses between 10 and 100 pC/$\mu m^2$ following the dose testing procedure as specified in supporting information note 3.

Transport and photovoltage experiments.
The measurements of temperature- and bias dependent resistivity are carried out in an AttoDry closed-cycle cryostat. Standard low-frequency lock-in techniques (Stanford Research Systems 860) and DC biasing (Yokogawa GS200 and Keithley 2400 SourceMeter) were used to measure the resistance and to bias the device. For the photovoltage experiments, a fixed frequency 1550nm laser was used to locally excite the device with a spatial resolution of ~2μm. Scanning mirrors and xyz-Piezo scanners were used to position the device and to scan the laser-spot across the device. The laser was modulated with an optical chopper and the photovoltage was measured with a lock-in detection on the chopper frequency. The incident laser power was adjusted using a variable optical attenuator (JGR OA5 l).

Continuous wave photomixing.
A fixed-frequency laser at 1550nm (193THz) and a tunable laser with 1528nm-1566nm (193THz+$\Delta f$) (ThorLabs C-band) are modulated at different chopping frequencies with a twin-slot chopper wheel (supporting information 8). The beams overlap in a beam splitter, which creates a sinusoidal optical beating at the difference frequency $\Delta f$. The sinusoidal detector response is detected with a lock-in amplifier (Stanford Research Systems 860) as a function of the difference frequency $\Delta f$ while demodulating at the difference frequency of the chopping frequencies.

# Supporting Information: A liquid nitrogen cooled superconducting transition edge sensor with ultra-high responsivity and GHz operation speeds

Paul Seifert[1,†], José Ramón Durán Retamal[1,†], Rafael Luque Merino[1], Hanan Herzig Sheinfux[1], John N. Moore[1], Mohammed Ali Aamir[1], Takashi Taniguchi[2], Kenji Wantanabe[3], Kazuo Kadowaki[4], Massimo Artiglia[5], Marco Romagnoli[5] and Dmitri K. Efetov[1]*

1. ICFO - Institut de Ciencies Fotoniques, The Barcelona Institute of Science and Technology, Castelldefels, Barcelona, 08860, Spain
2. International Center for Materials Nanoarchitectonics, National Institute for Materials Science, 1-1 Namiki, Tsukuba 305-0044, Japan
3. Research Center for Functional Materials, National Institute for Materials Science, 1-1 Namiki, Tsukuba 305-0044, Japan
4. Algae Biomass and Energy System (ABES) Research & Development Center & Institute of Materials Science University of Tsukuba 1-1-1, Tennodai, Tsukuba-shi, 305-8572, Japan
5. CNIT - National Interuniversity Consortium for Telecommunications, TeCIP -Institute of Communication, Information and Perception Technologies - Scuola Superiore Sant'Anna via Moruzzi, 1, 56124 Pisa, Italy

* E-mail: dmitri.efetov@icfo.eu
† equal contribution

**Contents:**

1. Van der Waals assembly of BSCCO nano-bolometer devices (S1)
2. Contact engineering (S2)
3. Focused helium ion nanofabrication:
    1. Dose testing (S3)
    2. Impact of ion irradiation and resolution (S4-S6)
4. Overview of BSCCO nano-bolometer devices and different bolometer designs (S7)
5. Channel homogeneity and notes on different bolometer designs (S8)
6. Hysteretic current-voltage relation as a function of temperature (S9)
7. Environmental stability of the BSCCO devices (S10)
8. Coupling efficiency at the grating coupler (S11)
9. Schematic of the continuous wave photomixing experiment (S12)

# 1. Van der Waals assembly of BSCCO nano-bolometer devices

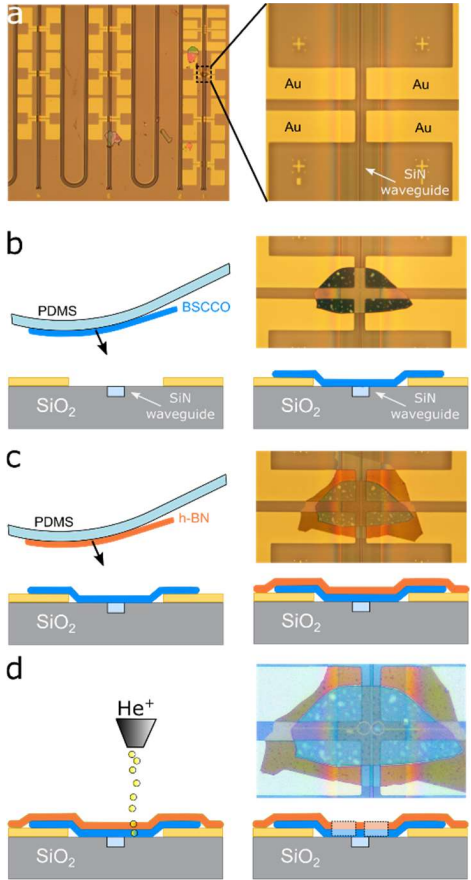

**Fig. S1. Schematic of the van der Waals assembly and helium FIB nano-fabrication. a,** Starting point is a SiO$_2$/SiN waveguide chip with pre-patterned electrodes close to the waveguide. **b,** A thin BSCCO crystal, which was micro-mechanically exfoliated via the scotch tape technique is transferred onto the waveguide and electrodes using a viscoelastic PDMS stamp on a glass substrate. **c,** In order to avoid degradation of the BSCCO, the flake is fully encapsulated with a thin crystal of h-BN, using the same process as for the BSCCO in (b). All exfoliation and van der Waals assembly steps are performed under inert Argon atmosphere inside a glovebox. **d,** A nanowire channel is patterned into the BSCCO on top of the SiN waveguide using focused helium ion beam irradiation.

# 2. Contact engineering

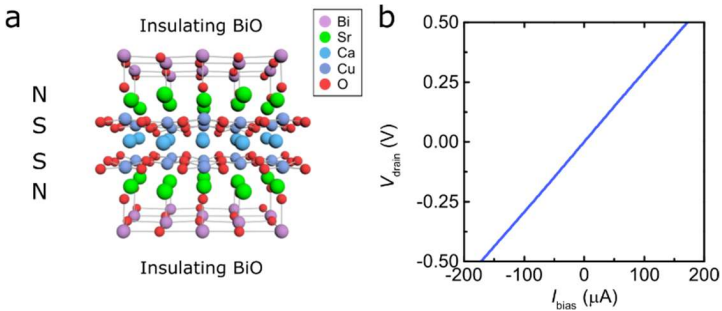

**Fig. S2. BSCCO layer structure and linear contact resistance. a,** Crystal structure - the outer bismuth oxide layers are considered to be insulating and the center layers of Sr and CuO$_2$ are described as a N-S-S-N structure (N – normal conductor, S - superconductor) **b,** Linear resistance in the normal state at 100K demonstrating a good electric contact through the insulating bismuth oxide layers.

**Contact engineering in BSCCO devices:** Figure S2 (a) depicts the layered structure of one BSCCO layer (half unit cell). The outermost bismuth oxide layers are considered to be insulating[1]. The strontium and copper oxide layers present a N-S-S-N structure where N denotes a normal conductor and S denotes a superconducting layer[1]. In order to make a reliable electrical contact to the superconducting layers in BSCCO one has to achieve a conducting path through the insulating BiO layer. This has been achieved in the past e.g. via in-situ selective ion etching through the top layer before metal evaporation[2] or via cold welding with diffusive metals e.g. indium[3]. The first approach is not promising for few unit cell sample thicknesses, because the BSCCO is subject to ambient conditions during nanofabrication. Cold-welding under inert atmosphere on the other hand gives good results even for half unit cell BSCCO layers. Here the electrical contact is obtained by contacting the bulk part of a BSCCO crystal which exhibits a monolayer thin region with indium[3]. Then, the bulk part of the BSCCO crystal can be patterned with a sharp AFM tip in order to isolate the monolayer region as the active area of the device[3]. The presence of the bulk flake however makes it difficult to encapsulate the whole BSCCO flake e.g. with h-BN to make it stable under ambient conditions. Hence, while this approach allows the study of monolayer BSCCO in fundamental science experiments, cold welding is not a favorable approach to reliably fabricate bolometer devices which need to withstand ambient conditions and multiple cooling cycles. For our nano-bolometer devices, we electrically contact the BSCCO crystals via a bottom contact using pre-fabricated gold electrodes. This allows us to fully encapsulate the BSCCO crystal with h-BN in order to make it stable under ambient conditions. In particular, we obtain the electrical contact to the superconducting layer by maximizing the effective contact interface between the BSCCO crystal and the electrodes. We find that for a contact interface > 200 $\mu m^2$ we get a reliable contact to the BSCCO crystal. While gold is not a diffusive metal as e.g. indium, the big contact interface makes it statistically favorable to contact the crystal via defects in the BiO plane or via diffusion of metal atoms. Figure S2 (b) shows a typical current-voltage relation of our devices after nano-fabrication of the BSCCO channel in the normal state (100 K). We find a good electrical contact to the device and a linear current-voltage characteristic. We note, that this method of contacting BSCCO is susceptible to a loss of electrical contact at low temperatures where roughly every second device loses contact to one of the electrodes during the first cooling cycle. However, this can be easily countered by the use of additional electrodes, as we find that electrodes which do not lose contact during the first cooling cycle are not likely to lose contact in subsequent cooling cycles. Overall, the possibility to make the devices environmentally stable makes our approach favorable for the use of few-layer BSCCO in applications.

## 3. Focused helium ion nanofabrication

### 3.1 Dose testing

**Dose testing for Helium ion nanofabrication**: In order to obtain the optimum helium ion dose for patterning nano-structures into BSCCO without a physical milling we utilize the following procedure: We irradiate an array of triangular shapes as indicated in Figure S3 with increasing dose on the BSCCO/h-BN stack. The triangle shapes are then imaged via the secondary electron generation from Helium ion imaging. In contrast to electron microscopy the secondary electrons are generated by positively charged Helium ions, which leads to an accumulation of positive charge on the sample. The BSCCO flake is grounded and hence allows for a depletion of the positive charge. For increasing ion dose of the patterned triangle shape, the BSCCO turns insulating which electrically isolates the pristine center of the BSCCO triangle from the surrounding grounded BSCCO. The accumulated positive charge cannot deplete from the isolated triangle anymore and the secondary electrons get trapped on the positively charged BSCCO which leads to a darkening of the isolated BSCCO in Helium ion imaging. The corresponding dose is used for patterning the nano-channel of the bolometer on the same flake.

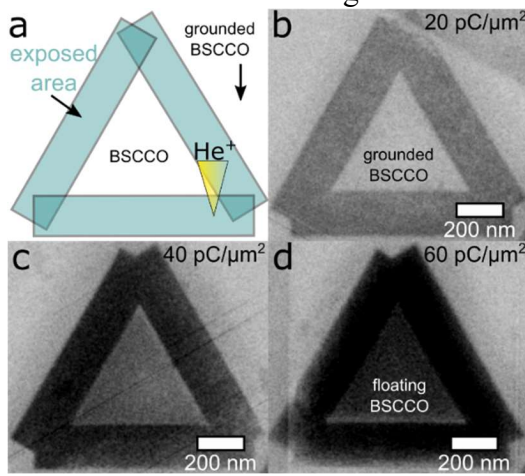

**Fig. S3. Helium ion dose testing. a,** Schematic illustration of the procedure for in-situ dose testing. The blue shaded triangular area is irradiated via a focused Helium ion beam. **b-d,** Secondary electron images from helium ion imaging of exposed BSCCO areas for increasing irradiation doses of 20 pC/μm², 40 pC/μm² and 60 pC/μm².

## 3.2 Impact of ion irradiation and resolution

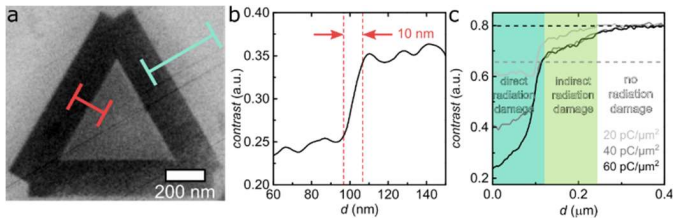

**Fig. S4. Resolution of He irradiation. a,** Secondary electron image from helium ion imaging of an exposed triangle with an ion dose of 40 pC/µm$^2$. **b,** Secondary electron imaging contrast along the red line in (a). **c,** Secondary electron imaging contrast along the blue line in (a) for irradiation doses of 20 pC/µm$^2$, 40 pC/µm$^2$ and 60 pC/µm$^2$.

**Resolution of nanofabrication and impact of helium ion irradiation:** Figure S4 (a) shows a secondary electron image from helium ion imaging of an exposed triangle with an ion dose of 40 pC/□m$^2$. From the line cut of the optical contrast along the red line (Fig. S4 (b)) we obtain a step edge between irradiated and non-irradiated area of ~10 nm. The step edge from the imaging contrast is a convolution of the step edge resolution and the imaging resolution from helium ion imaging. We estimate the nanofabrication to have a resolution of ~5 nm, which can be further improved via a more laborious alignment of the helium beam before patterning. Figure S4 (c) shows imaging contrast line cuts along the blue line in figure S4 (a) for different radiation doses. In addition to the direct radiation damage, we observe an indirect contrast change of the surrounding area with an extension of roughly 100 nm, which can also be seen as the darkened area around the exposed triangle in figure S4 (a). We attribute this to indirect irradiation damage as a result of backscattered helium ions which also expose the surrounding area with a reduced dose. Figure S5 shows the AFM topography of the patterned triangle as in Figs. S3 and S4. The helium ion irradiation can be associated with a height increase of the irradiated area, as observed in the topography map. The height increase of 1 nm in Fig. S5 (b) can be attributed to the accumulation of helium in the substrate during helium ion exposure which leads to the formation of helium bubbles [4,5] (see also following discussion on the simulated interaction of helium ions with an h-BN/BSCCO/SiO$_2$ device). As both the device topography and the imaging contrast in FIB are much sharper than the indirect irradiation damage, which we attribute to backscattered ions, the latter is the limiting factor in our nanofabrication. For the patterning of the BSCCO nano-wires, we try to account for the indirect irradiation damage and choose a width that leaves us with a channel of pristine BSCCO without radiation damage in the center, e.g. by choosing a channel width of 300nm to obtain a nominal ~100 nm wide channel of pristine BSCCO. We will later show, that this strategy works for short channels on the order of ~1□m, whereas the statistical occurrence of disorder/defects/bubbles imposes additional limitations on long nanowires. We note, that in future experiments, the damage from backscattered ions as well as the topography change can be addressed by the use of free-standing devices.

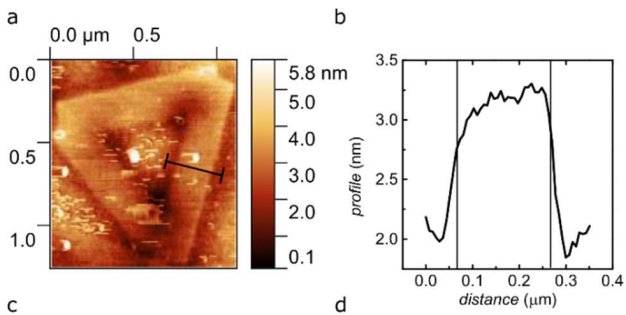

**Fig. S5. Sample topography after irradiation** AFM measurements on helium irradiated BSCCO/hBN flakes. **a, b** AFM topography map of the triangle shaped dose test in Fig S2. The helium ion irradiation leads to a height increase of the irradiated area as compared to the non-irradiated part. **b,** Line cut across the 200 nm wide pattern along the black line in (a). The height is increased by ~1nm.

**Simulation of helium ion interaction:** Figure S6 shows a Monte-Carlo simulation of the interaction of 30 keV helium ions with a typical device consisting of 20 nm h-BN / 9 nm BSCCO / 295 nm SiO$_2$ /Si. The simulation is performed using SRIM (**S**topping and **R**ange of **I**ons in **M**atter). Figure S6 (a) displays the ion trajectories of $10^3$ helium ions which enter perpendicularly at a single point into the device. The profile of the helium beam is almost unchanged after transmission through the hBN. After entering the $SiO_2$, the helium beam is increasingly deflected and the ion trajectory forms a pear shape below the point of entry. Figures S6 (b) and (c) show a cross section of the irradiation induced atomic displacement (b) and the distribution of stopped helium ions in the device after interaction of $10^4$ He ions (c). We find that the number of atomic displacements, which lead to defect formation and amorphization, concentrates in the BSCCO layer with a very sharp distribution in lateral dimensions. For increasing penetration depth the total number of atomic displacements increases but significantly broadens in close correspondence to the ion trajectories in figure S6 (a). Importantly, after primary interaction with the BSCCO device, most of the helium atoms accumulate in the $SiO_2$ substrate with an average ion range of 234 nm where the helium leads to a bubble formation and subsequent topography changes in our device.

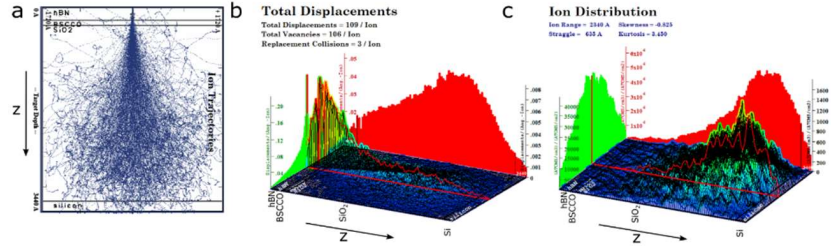

**Fig. S6. Simulation of helium ion interaction.** Monte-Carlo simulation of the interaction of helium ions with a h-BN/BSCCO/$SiO_2$ device using SRIM. **a,** Ion trajectories of $10^3$ helium ions with 30 keV which enter a typical stack of 20 nm h-BN / 9 nm BSCCO / 295 nm $SiO_2$ /Si. **b,** Displacement of atoms due to the interaction with $10^4$ helium ions. **c,** Distribution of stopped helium ions in the device after interacting with the stack. The average range of helium ions is 234 nm.

## 4. Overview of BSCCO nano-bolometer devices and different bolometer designs

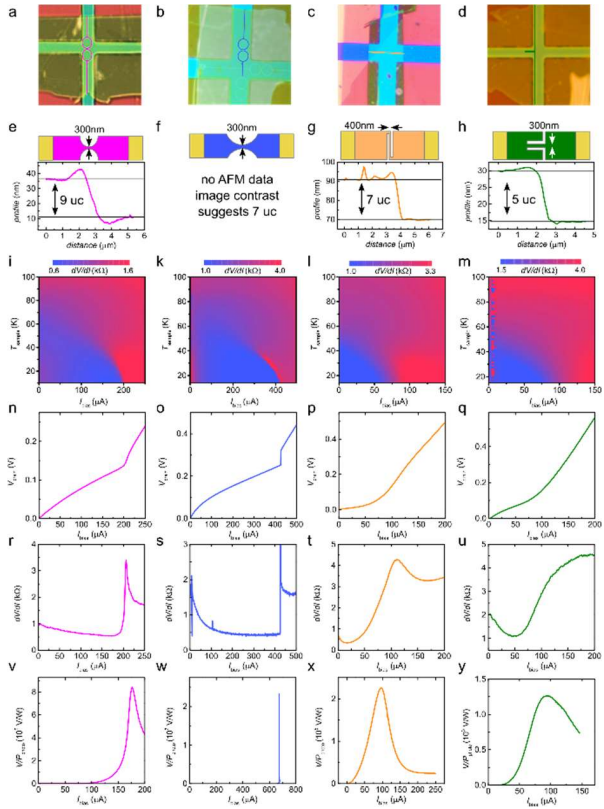

**Fig. S7. Overview of BSCCO nano-bolometer devices and different bolometer designs.** Overview of BSCCO nano-bolometer devices and different bolometer designs. The data in each column corresponds to one bolometer device. **a-d,** Optical microscopy images of BSCCO bolometer devices with different nano-constriction designs, as indicated by the colored lines. **e-h,** Bolometer designs and AFM profile of the bolometer devices. **i-m,** Two-terminal differential resistance d$V$/d$I$ of the devices as a function of bias current and device temperature. **n-p,** Current-voltage characteristic $IV$ of the bolometer devices at device temperatures of (n-10K, o-15K, p-15K, q-14K). **r-u**, Two-terminal differential resistance dV/dI of the devices as a function of bias current at device temperatures of (r-10K, s-15K, t-15K, u-14K). **v-y,** Voltage responsivity of the bolometer devices as a function of bias current at device temperatures of (v-10K, w-15K, x-15K, y-14K).

## 5. Channel homogeneity and notes on different bolometer designs

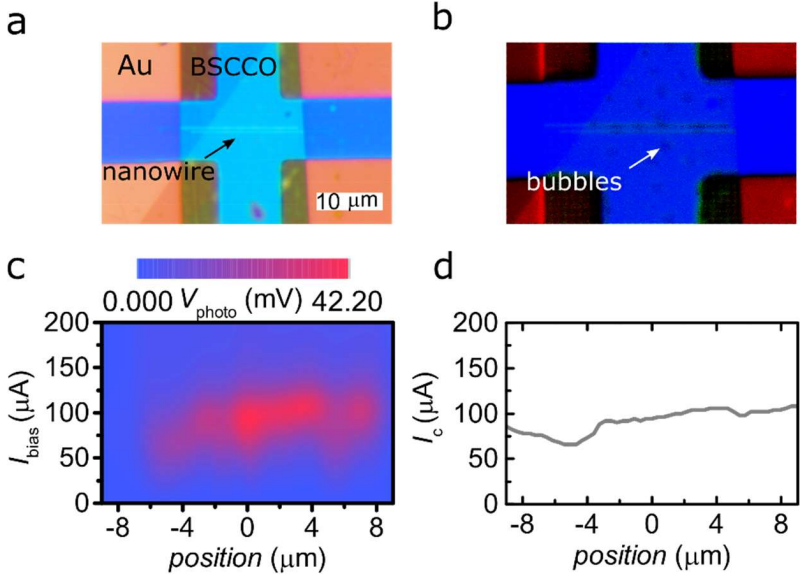

**Fig. S8. Channel homogeneity of the BSCCO nano-wires. a,** Optical image of the BSCCO nano-wire as in Fig. S7c comprising a 15µm long 200nm wide channel. **b,** Contrast enhanced optical microscopy image of the same device. Due to the stacking process of BSCCO and hBN bubbles are trapped at the position of the channel. **c,** Photo-response of the device along the channel as a function of the bias current. The response is inhomogeneous along the channel. **d,** Current of maximum photoresponse as a function of position in the nano-channel. We observe variations in the current of up to ~30%.

**Different BSCCO nano-bolometers and designs:** Figure S7 shows an overview of different BSCCO nano-bolometer devices and different bolometer designs on a Si/SiO$_2$ substrate. The data in each column corresponds to one bolometer device. The devices exhibit a BSCCO thickness between 5 and 9 unit cells. For the bolometer designs we chose both round channels, defined between two circle-shaped FIB lines ((a) and (b)) as well as straight nanowire like channels ((c) and (d)). All devices show superconducting behavior and a bolometric response. While devices (a), (c) and (d) showed only a small hysteresis in the current voltage relation (*IV*), indicative of a small contribution from self-heating, device (b) showed a very pronounced hysteresis loop (see also Fig. S9) and a detector latching up to 30 K. Correspondingly, the bolometric response is measured across a 1kΩ shunt resistor to avoid the latching in device (b). We observe, that the superconductivity, as visible in the differential resistance vs temperature (Fig. S7 (i)-(m)), is much sharper defined for the devices with a round nano-constriction ((a) and (b)) and enter superconductivity at a higher temperature as compared to the straight channel devices ((c) and (d)). This behavior is also reflected in the sharpness of the voltage response vs bias current (Fig. S7 (v)-(y)) and can be explained by the presence of disorder and/or bubbles in the BSCCO/h-BN devices. Figure S8 shows a corresponding analysis of device (c) comprising a 15 µm long 200nm wide nano-channel. In particular, this channel is much longer than the size of our laser-spot (~2-3 µm) and thus allows us to extract spatial information from the channel. In the contrast enhanced microscope image (b) one can already observe the occurrence of bubbles in the BSCCO/h-BN stack, which are distributed over the whole device and are also present at the position of the nano-channel. Figure S8 (c) shows the photovoltage response of the device along the channel as a function of the bias current. We observe, that the response occurs inhomogeneously along the channel and the maximum photoresponse occurs at different bias currents. Figure S8 (d) shows the corresponding current of maximum photoresponse along the channel. We observe variations in the current of up to ~30% as function of position in the nano-channel. We explain this by the occurrence of strain and disorder due to the bubble formation close to the channel. In addition, the presence of bubbles most likely influences the resolution and spatial homogeneity of the nanofabrication via helium ion beam irradiation. Devices comprising a round nano-channel have a well-defined weak link at the thinnest position of the channel. In this case, the presence of such inhomogeneity will lead to a different effective electronic width of the channel (compare critical current in devices as in Fig. S7 (a) and (b)). However, this will not smear out the superconducting transition as for a long channel (Fig. S7 (c) and (d)), where multiple superconducting transitions with different critical temperature and critical current can occur along the length of the channel.

## 6. Hysteretic current-voltage relation as a function of temperature

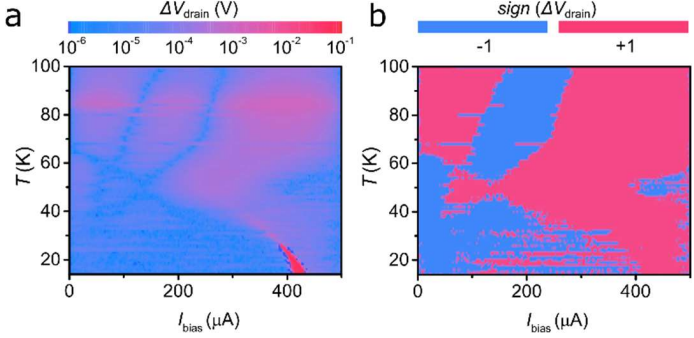

**Fig. S9. Hysteretic current-voltage relation.** Hysteresis of the superconducting transition in a nominal 100 nm nano-constriction (Fig. S7 (b)) as a function of temperature. **a,** Amplitude of the voltage difference between current sweeping directions as a function of bias current and temperature. **b,** Sign of the voltage difference between current sweeping directions as a function of bias current and temperature.

## 7. Environmental stability of the BSCCO devices

In previous work, the chemical instability of BSCCO, attributed to surface degradation and oxygen out-diffusion in ambient conditions and during thermocycling, was considered a bottleneck for the applicability of thin nanoscale high-Tc compounds in applications[3,6,7]. In order to avoid such degradation issues, we perform our device fabrication under inert argon atmosphere inside a glovebox. In particular, we fully encapsulte the BSCCO flakes with h-BN during our fabrication before we remove the devices from the glovebox for further fabrication. In addition, we utilize nanofabrication via focused helium ion beam irradiation to fabricate nanostructures in the BSCCO without a physical milling of the device such that the BSCCO stays protected from the environment via the h-BN encapsulation. Figure S10 shows the two-terminal differential resistance d$V$/d$I$ and current voltage relation $IV$ of two devices more that 30 and 50 days after device fabrication after more than one cooling cycle. We see, that the superconducting properties of the devices stay almost unchanged over prolonged periods of time and over multiple cooling cycles. In particular, the devices were subject to ambient conditions on many occasions after fabrication e.g. for wire-bonding and transport between laboratories and in the case of the device (a) and (b) even overnight, with no significant impact on the device properties.

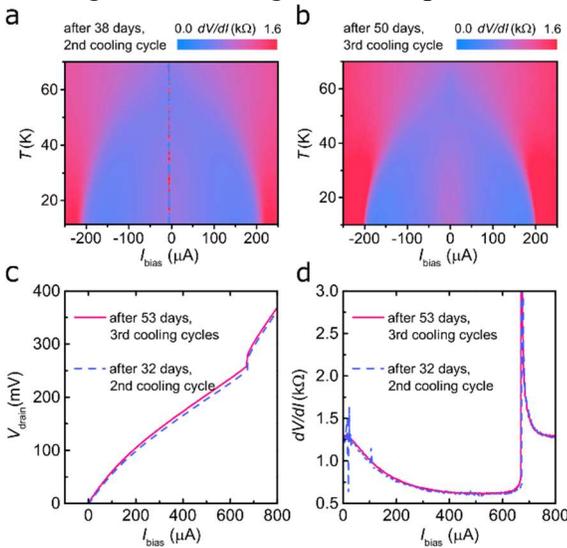

**Fig. S10. Environmental and thermocycling stability of the bolometer devices. a,** Two-terminal differential resistance d$V$/d$I$ of the device as in Fig. S7a as a function of bias current and device temperature after 38 days from fabrication in the 2$^{nd}$ cooling cycle. **b,** Two-terminal resistance d$V$/d$I$ of the same device after 50 days in the 3$^{rd}$ cooling cycle. The device was subject to ambient conditions before and between the cooling cycles. **c** and **d,** Current voltage characteristic $IV$ and differential resistance d$V$/d$I$ as a function of bias current at T = 15 K for the device as in Fig. S7b with 1kΩ parallel resistance after 32 days in the 2$^{nd}$ cooling cycle (blue dashed line) and after 53 days in the 3$^{rd}$ cooling cycle (pink line).

## 8. Coupling efficiency at the grating coupler

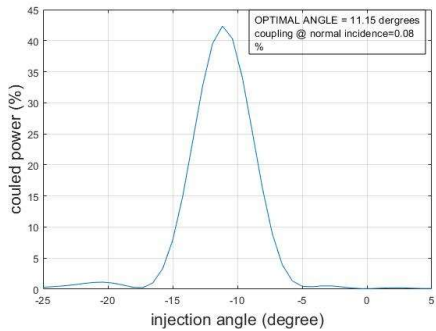

**Fig. S11. Simulation of coupling efficiency at the grating coupler.** Numerical simulation of the grating coupler's coupling efficiency into the SiN waveguide using our objective. The coupling efficiency exhibits a minimum at normal incidence, where we find a coupling efficiency of 0.08%.

## 9. Schematic of the continuous wave photomixing experiment

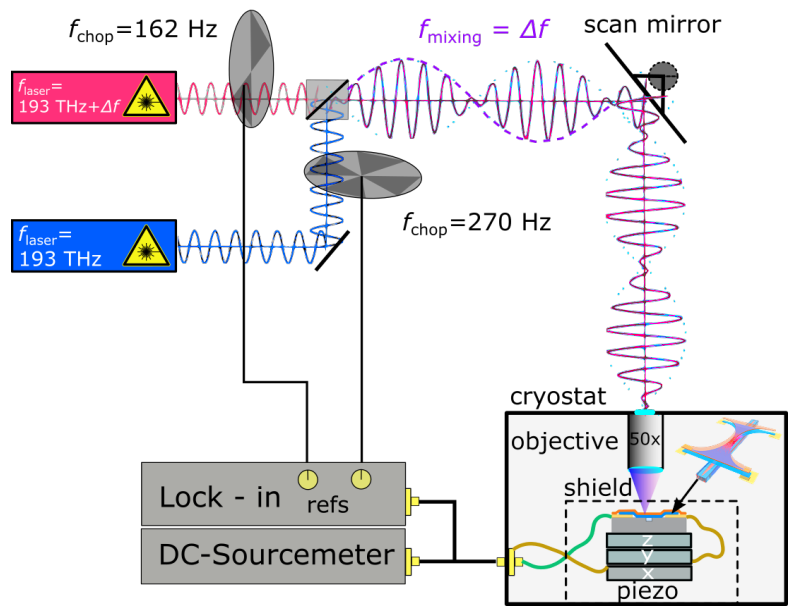

**Fig. S12. Schematic of the continuous wave photomixing experiment.** A fixed-frequency laser at 1550 nm (193 THz, blue) and a tunable laser with 1528 nm-1566 nm (193THz+$\Delta f$, pink) are modulated at different chopping frequencies with a twin-slot chopper wheel. The beams overlap in a beam splitter, which creates a sinusoidal optical beating at the mixing frequency $\Delta f$. Using a scanning mirror, the oscillating laser is focused onto the device, which is located on a xyz-piezo stage inside an AttoDry closed cycle cryostat. The bolometer is biased close to the critical current using a DC-sourcemeter and the sinusoidal bolometer response is detected with a lock-in amplifier at the difference frequency of the chopping frequencies. For a $\Delta f \leq 1/\tau$, the temperature can always follow the optical beating and the AC voltage response exhibits a maximum. For increasing mixing frequency $\Delta f \geq 1/\tau$, the bolometric response cannot follow the optical beating and the voltage response drops.